\pdfoutput=1
\documentclass[11pt]{article}
\usepackage[table,dvipsnames]{xcolor}
\usepackage{acl}
\usepackage{times}
\usepackage{latexsym}
\usepackage[T1]{fontenc}
\usepackage[utf8]{inputenc}
\usepackage{microtype}
\usepackage{inconsolata}
\usepackage{amsmath}
\usepackage{amssymb}
\usepackage{graphicx}
\usepackage{enumerate}
\usepackage{balance}
\usepackage{blindtext}
\usepackage{listings}
\usepackage{booktabs}
\usepackage{colortbl}
\usepackage{tcolorbox}
\usepackage{marvosym}
\usepackage{float}
\usepackage{makecell}
\usepackage{multicol}
\usepackage{multirow}
\usepackage{tabularx}
\usepackage{marvosym}
\usepackage{stfloats}
\usepackage{graphicx}
\usepackage{subfigure}
\usepackage{subcaption}
\usepackage{ulem}
\usepackage{hyperref}

\newcommand\footnoteONLYtext[1]{
    \let \mybackup \thefootnote
    \let \thefootnote \relax
    \footnotetext{#1}
    \let \thefootnote \mybackup
    \let \mybackup \imareallyundefinedcommand}

\lstdefinestyle{PythonCode}{
    language=Python,
    basicstyle=\ttfamily,
    breaklines=true,
    keywordstyle=\bfseries\color{NavyBlue},
    morekeywords={},
    emph={self},
    emphstyle=\bfseries\color{Rhodamine},
    commentstyle=\itshape\color{black!50!white},
    stringstyle=\bfseries\color{PineGreen!90!black},
    columns=flexible,
}

\lstdefinestyle{BashCode}{
    language=Bash,
    basicstyle=\ttfamily\color{white},
    backgroundcolor=\color{black},
    breaklines=true,
    keywordstyle=\bfseries\color{MidnightBlue},
    morekeywords={},
    emph={},
    emphstyle=\bfseries\color{Purple},
    commentstyle=\itshape\color{black!50!white},
    stringstyle=\bfseries\color{OliveGreen!90!black},
    columns=flexible,
}

\newcommand{\ie}{\textit{i.e., }}
\newcommand{\eg}{\textit{e.g., }}

\title{ChatDev: Communicative Agents for Software Development}

\author{
  \textbf{Chen Qian}{\footnotesize $^\bigstar$} \quad 
  \textbf{Wei Liu}{\footnotesize $^\bigstar$} \quad 
  \textbf{Hongzhang Liu}{\footnotesize $^\spadesuit$} \quad 
  \textbf{Nuo Chen}{\footnotesize $^\bigstar$} \quad 
  \textbf{Yufan Dang}{\footnotesize $^\bigstar$} \\
  \textbf{Jiahao Li}{\footnotesize $^\bigstar$} \quad 
  \textbf{Cheng Yang}{\footnotesize $^\clubsuit$} \quad 
  \textbf{Weize Chen}{\footnotesize $^\bigstar$} \quad 
  \textbf{Yusheng Su}{\footnotesize $^\bigstar$} \quad 
  \textbf{Xin Cong}{\footnotesize $^\bigstar$} \\
  \textbf{Juyuan Xu}{\footnotesize $^{\bigstar}$} \quad 
  \textbf{Dahai Li}{\footnotesize $^{\blacklozenge}$} \quad 
  \textbf{Zhiyuan Liu}{\footnotesize $^{\bigstar}$\textsuperscript{\Letter}} \quad  
  \textbf{Maosong Sun}{\footnotesize $^{\bigstar}$\textsuperscript{\Letter}} \\
  {\footnotesize $^\bigstar$}Tsinghua University \quad
  {\footnotesize $^\spadesuit$}The University of Sydney \quad 
  {\footnotesize $^\clubsuit$}BUPT \quad 
  {\footnotesize $^\blacklozenge$}Modelbest Inc.\\
  \href{mailto:qianc62@gmail.com}{\texttt{qianc62@gmail.com}} \quad 
  \href{mailto:liuzy@tsinghua.edu.cn}{\texttt{liuzy@tsinghua.edu.cn}} \quad 
  \href{mailto:sms@tsinghua.edu.cn}{\texttt{sms@tsinghua.edu.cn}}
}

\begin{document}

\maketitle
\footnoteONLYtext{\Letter: Corresponding Author.}

\begin{abstract}
Software development is a complex task that necessitates cooperation among multiple members with diverse skills.
Numerous studies used deep learning to improve specific phases in a waterfall model, such as design, coding, and testing.
However, the deep learning model in each phase requires unique designs, leading to technical inconsistencies across various phases, which results in a fragmented and ineffective development process.
In this paper, we introduce ChatDev, a chat-powered software development framework in which specialized agents driven by large language models (LLMs) are guided in what to communicate (via \textit{chat chain}) and how to communicate (via \textit{communicative dehallucination}). These agents actively contribute to the design, coding, and testing phases through unified language-based communication, with solutions derived from their multi-turn dialogues.
We found their utilization of natural language is advantageous for system design, and communicating in programming language proves helpful in debugging.
This paradigm demonstrates how linguistic communication facilitates multi-agent collaboration, establishing language as a unifying bridge for autonomous task-solving among LLM agents.
The code and data are available at \url{https://github.com/OpenBMB/ChatDev}.
\end{abstract}

\begin{figure}[t]
  \centering
  \includegraphics[width=0.50\textwidth]{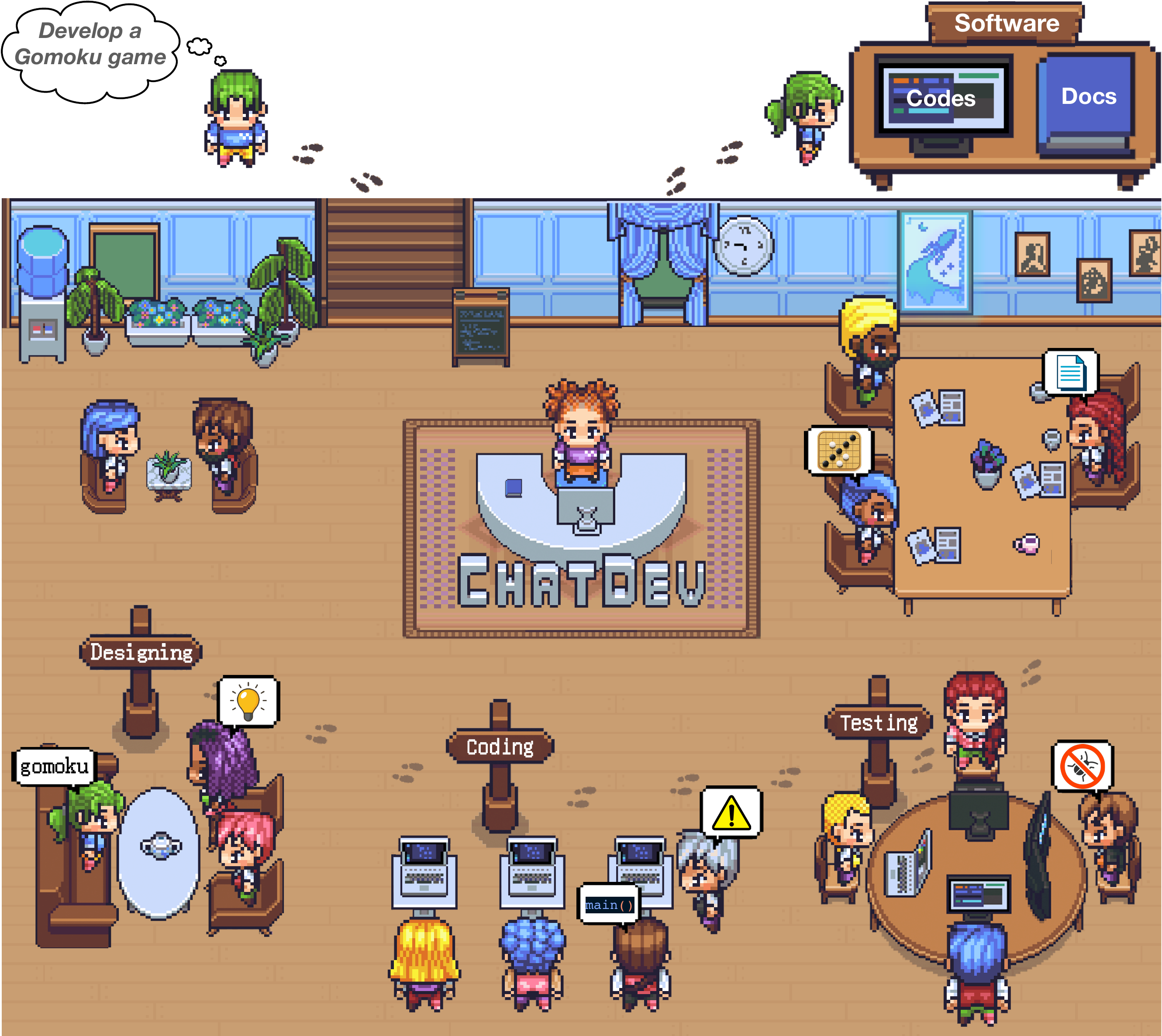}
  \caption{ChatDev, a \uwave{chat}-powered software \uwave{dev}elopment framework, integrates LLM agents with various social roles, working autonomously to develop comprehensive solutions via multi-agent collaboration.}
  \label{fig:chatdev}
\end{figure}

\section{Introduction}
Large language models (LLMs) have led to substantial transformations due to their ability to effortlessly integrate extensive knowledge expressed in language~\cite{NEURIPS2020_1457c0d6,bubeck2023sparks}, combined with their strong capacity for role-playing within designated roles~\cite{park2023generative,hua2023war,chen2023agentverse}.
This advancement eliminates the need for model-specific designs and delivers impressive performance in diverse downstream applications.
Furthermore, autonomous agents~\cite{AutoGPT,zhou2023webarena} have gained attention for enhancing the capabilities of LLMs with advanced features such as context-aware memory~\cite{sumers2023cognitive}, multi-step planning~\cite{liu2023bolaa}, and strategic tool using~\cite{schick2023toolformer}.

Software development is a complex task that necessitates cooperation among multiple members with diverse skills (\eg architects, programmers, and testers)~\cite{basili1989software,sawyer1998software}. 
This entails extensive communication among different roles to understand and analyze requirements through natural language, while also encompassing development and debugging using programming languages~\cite{ernst:LIPIcs.SNAPL.2017.4,banker1998software}.
Numerous studies use deep learning to improve specific phases of the waterfall model in software development, such as design, coding, and testing~\cite{DBLP:conf/re/PudlitzBV19,DBLP:journals/jss/MartinA15,DBLP:conf/wcre/GaoCXMSL19,DBLP:conf/icse/WangLT16}.
Due to these technical inconsistencies, methods employed in different phases remain isolated until now.
Every phase, from data collection and labeling to model training and inference, requires its unique designs, leading to a fragmented and less efficient development process in the field~\cite{DBLP:conf/icse/FreemanBSSDT01,ernst:LIPIcs.SNAPL.2017.4,DBLP:conf/se/WinklerGV20}.

Motivated by the expert-like potential of autonomous agents, we aim to establish language as a unifying bridge—utilizing multiple LLM-powered agents with specialized roles for cooperative software development through language-based communication across different phases; solutions in different phases are derived from their multi-turn dialogues, whether dealing with text or code.
Nevertheless, due to the tendency of LLM hallucinations~\cite{dhuliawala2023chain,zhang2023siren}, the strategy of generating software through communicative agents could lead to the non-trivial challenge of \textit{coding hallucinations}, which involves the generation of source code that is incomplete, unexecutable, or inaccurate, ultimately failing to fulfill the intended requirements~\cite{agnihotri2020systematic}.
The frequent occurrence of coding hallucination in turn reflects the constrained autonomy of agents in task completion, inevitably demanding additional manual intervention and thereby hindering the immediate usability and reliability of the generated software~\cite{ji2023survey}.

In this paper, we propose ChatDev (see Figure~\ref{fig:chatdev}), a \uwave{chat}-powered software-\uwave{dev}elopment framework integrating multiple "software agents" for active involvement in three core phases of the software lifecycle: design, coding, and testing.
Technically, ChatDev uses a \textit{chat chain} to divide each phase into smaller subtasks further, enabling agents' multi-turn communications to cooperatively propose and develop solutions (\eg creative ideas or source code).
The chain-structured workflow guides agents on what to communicate, fostering cooperation and smoothly linking natural- and programming-language subtasks to propel problem-solving.
Additionally, to minimize coding hallucinations, ChatDev includes an \textit{communicative dehallucination} mechanism, enabling agents to actively request more specific details before giving direct responses.
The communication pattern instructs agents on how to communicate, enabling precise information exchange for effective solution optimization while reducing coding hallucinations.
We built a comprehensive dataset containing software requirement descriptions and conducted comprehensive analyses.
The results indicate that ChatDev notably improves the quality of software, leading to improved completeness, executability, and better consistency with requirements.
Further investigations reveal that natural-language communications contribute to comprehensive system design, while programming-language communications drive software optimization.
In summary, the proposed paradigm demonstrates how linguistic communication facilitates multi-agent collaboration, establishing language as a unifying bridge for autonomous task-solving among LLM agents.

\section{Related Work}
Trained on vast datasets to comprehend and manipulate billions of parameters, LLMs have become pivotal in natural language processing due to their seamless integration of extensive knowledge~\cite{NEURIPS2020_1457c0d6,bubeck2023sparks,vaswani2017attention,radford2019language,touvron2023llama,wei2022emergent,Shanahan2023,chen2021evaluating,brants2007large,chen2021evaluating,ouyang2022training,yang2023large,qin2023large,kaplan2020scaling}.
Furthermore, LLMs have demonstrated strong role-playing abilities~\cite{li2023camel,park2023generative,hua2023war,chan2023chateval,zhou2023agents,chen2023agentverse,chen2023gamegpt,Cohen2023LMVL,Li2023MetaAgentsSI}. 
Recent progress, particularly in the field of autonomous agents~\cite{zhou2023webarena,wang2023voyager,park2023generative,wang2023humanoid,AutoGPT,GPTEngineer,wang2023promptagent}, is largely attributed to the foundational advances in LLMs. These agents utilize the robust capabilities of LLMs, displaying remarkable skills in memory~\cite{park2023generative,sumers2023cognitive}, planning~\cite{chen2023agentverse,liu2023bolaa} and tool use~\cite{schick2023toolformer,cai2023large,qin2023toolllm,ruan2023tptu,GPT4Tools}, enabling them to reason in complex scenarios~\cite{wei2022chain,zhao2023expel,zhou2023webarena,ma2023laser,zhang2023generative,wang2023large,ding2023designgpt,weng2023prompt}.

Software development is a multifaceted and intricate process that requires the cooperation of multiple experts from various fields~\cite{yilmaz2012systematic,acuna2006emphasizing,basili1989software,sawyer1998software,banker1998software,france2007model}, encompassing the requirement analysis and system design in natural languages~\cite{DBLP:conf/re/PudlitzBV19,DBLP:journals/jss/MartinA15,DBLP:conf/icse/NaharZLK22}, along with system development and debugging in programming languages~\cite{DBLP:conf/wcre/GaoCXMSL19,DBLP:conf/icse/WangLT16,DBLP:conf/icse/WanLXLHM022}.
Numerous studies employ the waterfall model, a particular software development life cycle, to segment the process into discrete phases (\eg design, coding, testing) and apply deep learning to improve the effectiveness of certain phases~\cite{DBLP:conf/se/WinklerGV20,DBLP:conf/icse/EzziniA0S22,DBLP:conf/wcre/ThallerLE19,DBLP:conf/icse/ZhaoCL021,DBLP:conf/iclr/NijkampPHTWZSX23,DBLP:conf/kbse/WanZYXY0Y18,DBLP:conf/icse/WangSSWWN21}. 

\section{ChatDev}
\begin{figure*}[htbp]
  \centering
  \includegraphics[width=0.88\textwidth]{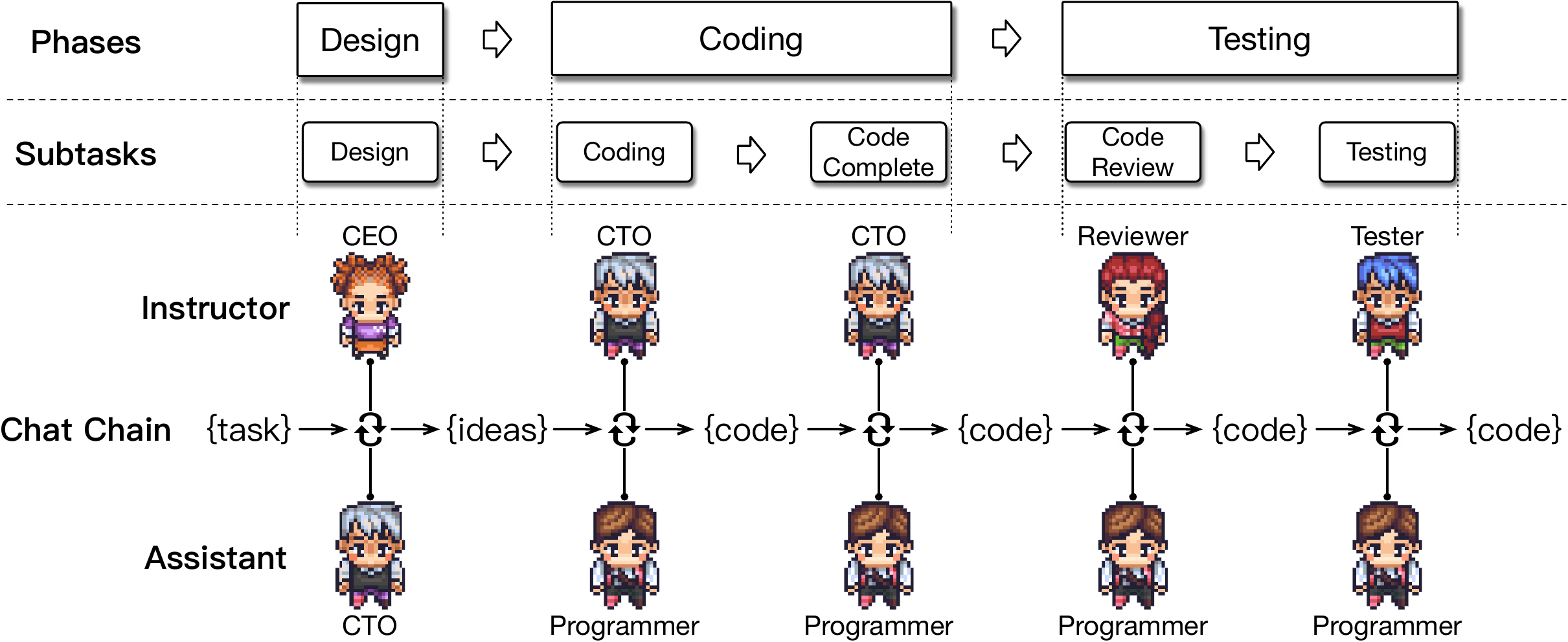}
  \caption{Upon receiving a preliminary task requirement (\eg ``\textit{develop a Gomoku game}''), these software agents engage in multi-turn communication and perform instruction-following along a chain-structured workflow, collaborating to execute a series of subtasks autonomously to craft a comprehensive solution.}
  \label{fig:chatchain}
\end{figure*}

We introduce ChatDev, a \uwave{chat}-powered software-\uwave{dev}elopment framework that integrates multiple "software agents" with various social roles (\eg requirements analysts, professional programmers and test engineers) collaborating in the core phases of the software life cycle, see Figure~\ref{fig:chatdev}.
Technically, to facilitate cooperative communication, ChatDev introduces \textit{chat chain} to further break down each phase into smaller and manageable subtasks, which guides multi-turn communications between different roles to propose and validate solutions for each subtask.
In addition, to alleviate unexpected hallucinations, a communicative pattern named \textit{communicative dehallucination} is devised, wherein agents request more detailed information before responding directly and then continue the next round of communication based on these details.

\subsection{Chat Chain}
Although LLMs show a good understanding of natural and programming languages, efficiently transforming textual requirements into functional software in a single step remains a significant challenge.
ChatDev thus adopts the core principles of the waterfall model, using a chat chain ($\mathcal{C}$) with sequential phases ($\mathcal{P}$), each comprising sequential subtasks ($\mathcal{T}$). Specifically, ChatDev segments the software development process into three sequential phases: design, coding, and testing.
The coding phase is further subdivided into subtasks of code writing and completion, and the testing phase is segmented into code review (static testing) and system testing (dynamic testing), as illustrated in Figure~\ref{fig:chatchain}.
In every subtask, two agents, each with their own specialized roles (\eg a reviewer skilled at identifying endless loops and a programmer adept in GUI design), perform the functions of an instructor ($\mathcal{I}$) and an assistant ($\mathcal{A}$).
The instructor agent initiates instructions, instructing ($\rightarrow$) the discourse toward the completion of the subtask, while the assistant agent adheres to these instructions and responds with ($\leadsto$) appropriate solutions.
They engage in a multi-turn dialogue ($\mathsf{C}$), working cooperatively until they achieve consensus, extracting ($\tau$) solutions that can range from the text (\eg defining a software function point) to code (\eg creating the initial version of source code), ultimately leading to the completion of the subtask.
The entire task-solving process along the agentic workflow can be formulated as:
\begin{equation}
\begin{aligned}
& \mathcal{C} = \langle \mathcal{P}^{1}, \mathcal{P}^{2} ,\dots,\mathcal{P}^{|\mathcal{C}|} \rangle \\
& \mathcal{P}^i = \langle \mathcal{T}^1, \mathcal{T}^2,\dots, \mathcal{T}^{|\mathcal{P}^i|} \rangle \\
& \mathcal{T}^j = \tau \big( \mathsf{C}(\mathcal{I},\mathcal{A}) \big) \\
& \mathsf{C}(\mathcal{I},\mathcal{A}) = {\langle \mathcal{I}\rightarrow\mathcal{A}, \ \mathcal{A}\leadsto\mathcal{I} \rangle}_{\circlearrowleft} \\
\end{aligned}
\end{equation}
The dual-agent communication design simplifies communications by avoiding complex multi-agent topologies, effectively streamlining the consensus-reaching process~\cite{yin-etal-2023-exchange,chen2023agentverse}.
Subsequently, the solutions from previous tasks serve as bridges to the next phase, allowing a smooth transition between subtasks. This approach continues until all subtasks are completed.
It's worth noting that the conceptually simple but empirically powerful chain-style structure guides agents on what to communicate, fostering cooperation and smoothly linking natural- and programming-language subtasks.
It also offers a transparent view of the entire software development process, allowing for the examination of intermediate solutions and assisting in identifying possible problems.

\paragraph{Agentization}
To enhance the quality and reduce human intervention, ChatDev implements prompt engineering that only takes place at the start of each subtask round. As soon as the communication phase begins, the instructor and the assistant will communicate with each other in an automated loop, continuing this exchange until the task concludes.
However, simply exchanging responses cannot achieve effective multi-round task-oriented communication, since it inevitably faces significant challenges including role flipping, instruction repeating, and fake replies.
As a result, there is a failure to advance the progression of productive communications and hinders the achievement of meaningful solutions.
ChatDev thus employs inception prompting mechanism~\cite{li2023camel} for initiating, sustaining, and concluding agents' communication to guarantee a robust and efficient workflow.
This mechanism is composed of the instructor system prompt $\mathsf{P}_I$ and the assistant system prompt $\mathsf{P}_A$.
The system prompts for both roles are mostly symmetrical, covering the overview and objectives of the current subtask, specialized roles, accessible external tools, communication protocols, termination conditions, and constraints or requirements to avoid undesirable behaviors.
Then, an instructor $\mathcal{I}$ and an assistant $\mathcal{A}$ are instantiated by hypnotizing LLMs via $\mathsf{P}_I$ and $\mathsf{P}_A$:
\begin{equation}
\begin{aligned}
\mathcal{I} = \rho(LLM, \mathsf{P}_I), \ \ \mathcal{A} = \rho(LLM, \mathsf{P}_A)
\end{aligned}
\end{equation}
where $\rho$ is the role customization operation, implemented via system message assignment.

\paragraph{Memory}
Note that the limited context length of common LLMs typically restricts the ability to maintain a complete communication history among all agents and phases.
To tackle this issue, based on the nature of the chat chain, we accordingly segment the agents' context memories based on their sequential phases, resulting in two functionally distinct types of memory: \textit{short-term memory} and \textit{long-term memory}.
Short-term memory is utilized to sustain the continuity of the dialogue within a single phase, while long-term memory is leveraged to preserve contextual awareness across different phases.

Formally, short-term memory records an agent's current phase utterances, aiding context-aware decision-making.
At the time $t$ during phase $\mathcal{P}^i$, we use $\mathcal{I}_t^i$ to represent the instructor's instruction and $\mathcal{A}_t^i$ for the assistant's response.
The short-term memory $\mathcal{M}$ collects utterances up to time $t$ as:
\begin{equation}
\mathcal{M}_t^i = \langle (\mathcal{I}_1^i,\mathcal{A}_1^i), (\mathcal{I}_2^i,\mathcal{A}_2^i), \ldots, (\mathcal{I}_t^i,\mathcal{A}_t^i) \rangle
\end{equation}

In the next time step $t+1$, the instructor utilizes the current memory to generate a new instruction $\mathcal{I}_{t+1}^i$, which is then conveyed to the assistant to produce a new response $\mathcal{A}_{t+1}^i$. The short-term memory iteratively updates until the number of communications reaches the upper limit $|\mathcal{M}^i|$:
\begin{equation}
\begin{aligned}
\mathcal{I}_{t+1}^i = &\mathcal{I}(\mathcal{M}_t^i), \ \ \mathcal{A}_{t+1}^i = \mathcal{A}(\mathcal{M}_t^i, \mathcal{I}_{t+1}^i) \\
&\mathcal{M}_{t+1}^i = \mathcal{M}_t^i \cup (\mathcal{I}_{t+1}^i, \mathcal{A}_{t+1}^i)
\end{aligned}
\end{equation}

To perceive dialogues through previous phases, the chat chain only transmits the solutions from previous phases as long-term memories $\tilde{\mathcal{M}}$, integrating them at the start of the next phase and enabling the cross-phase transmission of long dialogues:
\begin{equation}
\mathcal{I}_1^{i+1} = \tilde{\mathcal{M}}^{i} \cup \mathsf{P}^{i+1}_{\mathcal{I}}, \ \ \tilde{\mathcal{M}}^{i} = \bigcup_{j=1}^{i} \tau(\mathcal{M}_{|\mathcal{M}^j|}^{j})
\end{equation}
where \(\mathsf{P}\) symbolizes a predetermined prompt that appears exclusively at the start of each phase.

By sharing only the solutions of each subtask rather than the entire communication history, ChatDev minimizes the risk of being overwhelmed by too much information, enhancing concentration on each task and encouraging more targeted cooperation, while simultaneously facilitating cross-phase context continuity.

\subsection{Communicative Dehallucination}
LLM hallucinations manifest when models generate outputs that are nonsensical, factually incorrect, or inaccurate~\cite{dhuliawala2023chain,zhang2023siren}. This issue is particularly concerning in software development, where programming languages demand precise syntax—the absence of even a single line can lead to system failure.
We have observed that LLMs often produce \textit{coding hallucinations}, which encompass potential issues like incomplete implementations, unexecutable code, and inconsistencies that don't meet requirements.
Coding hallucinations frequently appear when the assistant struggles to precisely follow instructions, often due to the vagueness and generality of certain instructions that require multiple adjustments, making it challenging for agents to achieve full compliance.
Inspired by this, we introduce \textit{communicative dehallucination}, which encourages the assistant to actively seek more detailed suggestions from the instructor before delivering a formal response.

Specifically, a vanilla communication pattern between the assistant and the instructor follows a straightforward instruction-response format:
\begin{equation}
\langle \mathcal{I}\rightarrow\mathcal{A}, \ \mathcal{A}\leadsto\mathcal{I} \rangle_{\circlearrowleft}
\end{equation}
In contrast, our communicative dehallucination mechanism features a deliberate "role reversal", where the assistant takes on an instructor-like role, proactively seeking more specific information (\eg the precise name of an external dependency and its related class) before delivering a conclusive response. After the instructor provides a specific modification suggestion, the assistant proceeds to perform precise optimization:
\begin{equation}
\langle \mathcal{I}\rightarrow\mathcal{A},\  \langle\mathcal{A}\rightarrow\mathcal{I},\ \mathcal{I}\leadsto\mathcal{A}\rangle_\circlearrowleft,\  \mathcal{A}\leadsto\mathcal{I} \rangle_\circlearrowleft
\end{equation}
Since this mechanism tackles one concrete issue at a time, it requires multiple rounds of communication to optimize various potential problems.
The communication pattern instructs agents on how to communicate, enabling finer-grained information exchange for effective solution optimization, which practically aids in reducing coding hallucinations.

\section{Evaluation}

\begin{table*}[h]
\centering
\begin{tabular}{lccccc}
\toprule[1.5pt]
\textbf{Method} & \textbf{Paradigm} & \textbf{Completeness} & \textbf{Executability} & \textbf{Consistency} & \textbf{Quality} \\
\midrule[0.75pt]
GPT-Engineer & \includegraphics[height=10pt]{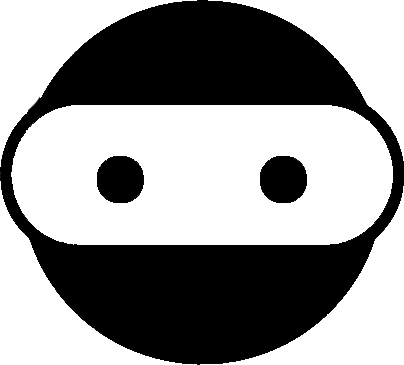} & \underline{0.5022}$^\dagger$ & 0.3583$^\dagger$ & \underline{0.7887}$^\dagger$ & 0.1419$^\dagger$ \\
MetaGPT & \includegraphics[height=10pt]{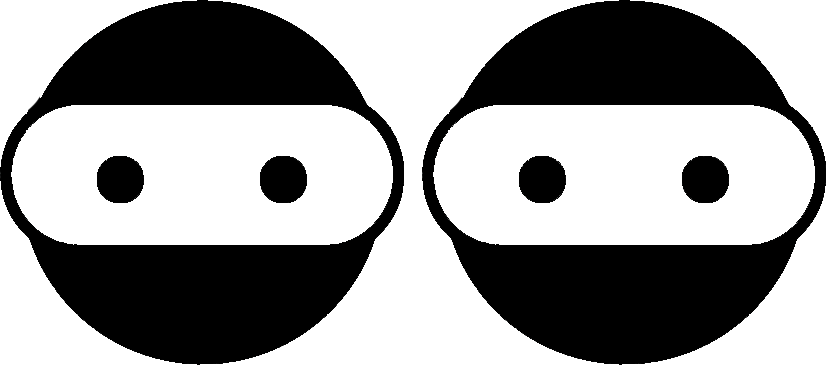} & 0.4834$^\dagger$ & \underline{0.4145}$^\dagger$ & 0.7601$^\dagger$ & \underline{0.1523}$^\dagger$ \\
ChatDev & \includegraphics[height=10pt]{figs/p5.png} & \textbf{0.5600} & \textbf{0.8800} & \textbf{0.8021} & \textbf{0.3953} \\
\bottomrule[1.5pt]
\end{tabular}
\caption{Overall performance of the LLM-powered software development methods, encompassing both single-agent (\includegraphics[height=8pt]{figs/p4.png}) and multi-agent (\includegraphics[height=8pt]{figs/p5.png}) paradigms. Performance metrics are averaged for all tasks. The top scores are in bold, with second-highest underlined. $\dagger$ indicates significant statistical differences (p$\le$0.05) between a baseline and ours.}
\label{tab:main-results}
\end{table*}

\begin{table}[t]
\centering
\resizebox{0.48\textwidth}{!}{
\begin{tabular}{lcccc}
\toprule[1.5pt]
\textbf{Method} & \textbf{Evaluator} & \textbf{Baseline Wins} & \textbf{ChatDev Wins} & \textbf{Draw} \\
\midrule[0.75pt]
\multirow{2}{*}{GPT-Engineer} & GPT-4 & 22.50\% & 77.08\% & 00.42\% \\
 & Human & 09.18\% & 90.16\% & 00.66\% \\
\midrule[0.25pt]
\multirow{2}{*}{MetaGPT}& GPT-4 & 37.50\% & 57.08\% & 05.42\% \\
 & Human & 07.92\% & 88.00\% & 04.08\% \\
\bottomrule[1.5pt]
\end{tabular}
}
\caption{Pairwise evaluation results.}
\label{tab:human_eval}
\end{table}

\noindent \paragraph{Baselines} We chose some representative LLM-based software development methods as our baselines.
GPT-Engineer~\cite{GPTEngineer} is a fundamental single-agent approach in LLM-driven software agents with a precise understanding of task requirements and the application of one-step reasoning, which highlights its efficiency in generating detailed software solutions at the repository level.
MetaGPT~\cite{hong2023metagpt} is an advanced framework that allocates specific roles to various LLM-driven software agents and incorporates standardized operating procedures to enable multi-agent participation. In each step agents with specific roles generate solutions by adhering to static instructions predefined by human experts.

\noindent \paragraph{Datasets} Note that, as of now, there isn't a publicly accessible dataset containing textual descriptions of software requirements in the context of agent-driven software development. 
To this end, we are actively working towards developing a comprehensive dataset for software requirement descriptions, which we refer to as SRDD (Software Requirement Description Dataset).
Drawing on previous work~\cite{li2023camel}, we utilize existing software descriptions as initial examples, which are then further developed through a process that combines LLM-based automatic generation with post-processing refinement guided by humans.
As a result, this dataset includes important software categories from popular platforms such as Ubuntu, Google Play, Microsoft Store, and Apple Store. It comprises 1,200 software task prompts that have been carefully categorized into 5 main areas: Education, Work, Life, Game, and Creation.
All these areas are further divided into 40 subcategories, and each subcategory contains 30 unique task prompts.

\noindent \paragraph{Metrics} Evaluating software is also a challenging task, especially when trying to assess it on a holistic level. 
Under the current limitation of scarce benchmark resources, traditional function-oriented code generation metrics (\eg \texttt{pass@k}), cannot seamlessly transfer to a comprehensive evaluation of entire software systems.
The main reason for this is that it is often impractical to develop manual or automated test cases for various types of software, especially those involving complex interfaces, frequent user interactions, or non-deterministic feedback.
As an initial strategy, we apply three fundamental and objective dimensions that reflect different aspects of coding hallucinations to evaluate the agent-generated software, and then integrate them to facilitate a more holistic evaluation:
\begin{enumerate}[$\bullet$]
\setlength{\topsep}{-4pt}
\setlength{\itemsep}{-4pt}
\item \textit{Completeness} measures the software's ability to fulfill code completion in software development, quantified as the percentage of software without any "placeholder" code snippets. A higher score indicates a higher probability of automated completion.
\item \textit{Executability} assesses the software's ability to run correctly within a compilation environment, quantified as the percentage of software that compiles successfully and can run directly. A higher score indicates a higher probability of successful execution.
\item \textit{Consistency} measures how closely the generated software code aligns with the original requirement description, quantified as the cosine distance between the semantic embeddings of the textual requirements and the generated software code\footnote{Comments should be excluded from the code to avoid potential information leakage during evaluations.}. A higher score indicates a greater degree of consistency with the requirements.
\item \textit{Quality} is a comprehensive metric that integrates various factors to assess the overall quality of software, quantified by multiplying\footnote{One can also choose to average the sub-metrics, which yields similar trends.} completeness, executability, and consistency. A higher quality score suggests a higher overall satisfaction with the software generated, implying a lower need for further manual intervention.
\end{enumerate}

\noindent \paragraph{Implementation Details} We divided software development into 5 subtasks within 3 phases, assigning specific roles like CEO, CTO, programmer, reviewer, and tester. A subtask would terminate and get a conclusion either after two unchanged code modifications or after 10 rounds of communication. During the code completion, review, and testing, a communicative dehallucination is activated. For ease of identifying solutions, the assistant begins responses with "<SOLUTION>" when a consensus is reached. We used ChatGPT-3.5 with a temperature of 0.2 and integrated Python-3.11.4 for feedback. All baselines in the evaluation share the same hyperparameters and settings for fairness.

\subsection{Overall Performance}
As illustrated in Table~\ref{tab:main-results}, ChatDev outperforms all baseline methods across all metrics, showing a considerable margin of improvement.
Firstly, the improvement of ChatDev and MetaGPT over GPT-Engineer demonstrates that complex tasks are difficult to solve in a single-step solution. Therefore, explicitly decomposing the difficult problem into several smaller, more manageable subtasks enhances the effectiveness of task completion.
Additionally, in comparison to MetaGPT, ChatDev significantly raises the \textit{Quality} from 0.1523 to 0.3953. This advancement is largely attributed to the agents employing a cooperative communication method, which involves autonomously proposing and continuously refining source code through a blend of natural and programming languages, as opposed to merely delivering responses based on human-predefined instructions.
The communicative agents guide each subtask towards integrated and automated solutions, efficiently overcoming the restrictions typically linked to manually established optimization rules, and offering a more versatile and adaptable framework for problem-solving.

\begin{table}[t]
\centering
\resizebox{0.48\textwidth}{!}{
\begin{tabular}{lcccc}
\toprule[1.5pt]
\textbf{Method} & \textbf{Duration} (s) & \textbf{\#Tokens} & \textbf{\#Files} & \textbf{\#Lines} \\
\midrule[0.75pt]
GPT-Engineer & 15.6000 & 7,182.5333 & 3.9475 & 70.2041 \\
MetaGPT & 154.0000 & 29,278.6510 & 4.4233 & 153.3000 \\
ChatDev & 148.2148 & 22,949.4450 & 4.3900 & 144.3450 \\
\bottomrule[1.5pt]
\end{tabular}
}
\caption{Software statistics include Duration (time consumed), \#Tokens (number of tokens used), \#Files (number of code files generated), and \#Lines (total lines of code across all files) in the software generation process.}
\label{tab:statistics}
\end{table}

To further understand user preferences in practical settings, we use the setting adopted by~\citet{li2023camel}, where agent-generated solutions are compared in pairs by both human participants and the prevalent GPT-4 model to identify the preferred one.\footnote{For fairness, GPT-4's evaluation mitigated possible positional bias~\cite{wang2023largelanguage}, and human experts independently assessed the task solutions, randomized to prevent order bias.} Table~\ref{tab:human_eval} shows ChatDev consistently outperforming other baselines, with higher average win rates in both GPT-4 and human evaluations.

Furthermore, the software statistics presented in Table~\ref{tab:statistics} indicates that the multi-agent paradigm, despite being slower and consuming more tokens than the single-agent method, yields a greater number of code files and a larger codebase, which may enhance the software's functionality and integrity.
Analyzing the dialogues of agents suggests that the multi-agent communication method often leads agents to autonomously offer functional enhancements (\eg GUI creation or increasing game difficulty), thereby potentially resulting in the incorporation of beneficial features that were not explicitly specified in requirements.
Taking all these factors together, we posit that the fundamental characteristics of multi-agent software development take on greater significance, surpassing short-term concerns like time and economic costs in the current landscape.

\subsection{Ablation Study}

\begin{table}[t]
\centering
\resizebox{0.48\textwidth}{!}{
\begin{tabular}{lcccc}
\toprule[1.5pt]
\textbf{Variant} & \textbf{Completeness} & \textbf{Executability} & \textbf{Consistency} & \textbf{Quality} \\
\midrule[0.75pt]
ChatDev & 0.5600 & \textbf{0.8800} & \textbf{0.8021} & \textbf{0.3953} \\
\midrule[0.25pt]
$\leq$Coding & 0.4100 & 0.7700 & 0.7958 & 0.2512 \\
$\leq$Complete & \textbf{0.6250} & 0.7400 & 0.7978 & 0.3690 \\
$\leq$Review & \underline{0.5750} & \underline{0.8100} & \underline{0.7980} & \underline{0.3717} \\
$\leq$Testing & 0.5600 & \textbf{0.8800} & \textbf{0.8021} & \textbf{0.3953} \\
\midrule[0.25pt]
$\diagdown$CDH & 0.4700 & 0.8400 & 0.7983 & 0.3094 \\
$\diagdown$Roles & 0.5400 & 0.5800 & 0.7385 & 0.2212 \\
\bottomrule[1.5pt]
\end{tabular}
}
\caption{Ablation study on main components or mechanisms. $\leq x$ denotes halting the chat chain after the completion of the $x$ phrase, and $\diagdown$ denotes the removing operation. CDH denotes the communicative dehallucination mechanism.}
\label{tab:ablation}
\end{table}

This section examines key components or mechanisms within our multi-agent cooperation framework by removing particular phases in the chat chain, communicative dehallucination, or the roles assigned to all agents in their system prompts.
Figure~\ref{tab:ablation} shows that the code complete phase enhances \textit{Completeness}, with testing critical for \textit{Executability}. 
\textit{Quality} steadily rises with each step, suggesting that software development optimization is progressively attained through multi-phase communications among intelligent agents.
Meanwhile, eliminating communicative dehallucination results in a decrease across all metrics, indicating its effectiveness in addressing coding hallucinations.
Most interestingly, the most substantial impact on performance occurs when the roles of all agents are removed from their system prompts.
Detailed dialogue analysis shows that assigning a "prefer GUI design" role to a programmer results in generated source code with relevant GUI implementations; in the absence of such role indications, it defaults to implement unfriend command-line-only programs only.
Likewise, assigning roles such as a "careful reviewer for bug detection" enhances the chances of discovering code vulnerabilities; without such roles, feedback tends to be high-level, leading to limited adjustments by the programmer.
This finding underscores the importance of assigning roles in eliciting responses from LLMs, underscoring the significant influence of multi-agent cooperation on software quality.

\begin{figure}[t]
    \centering
    \includegraphics[width=0.40\textwidth]{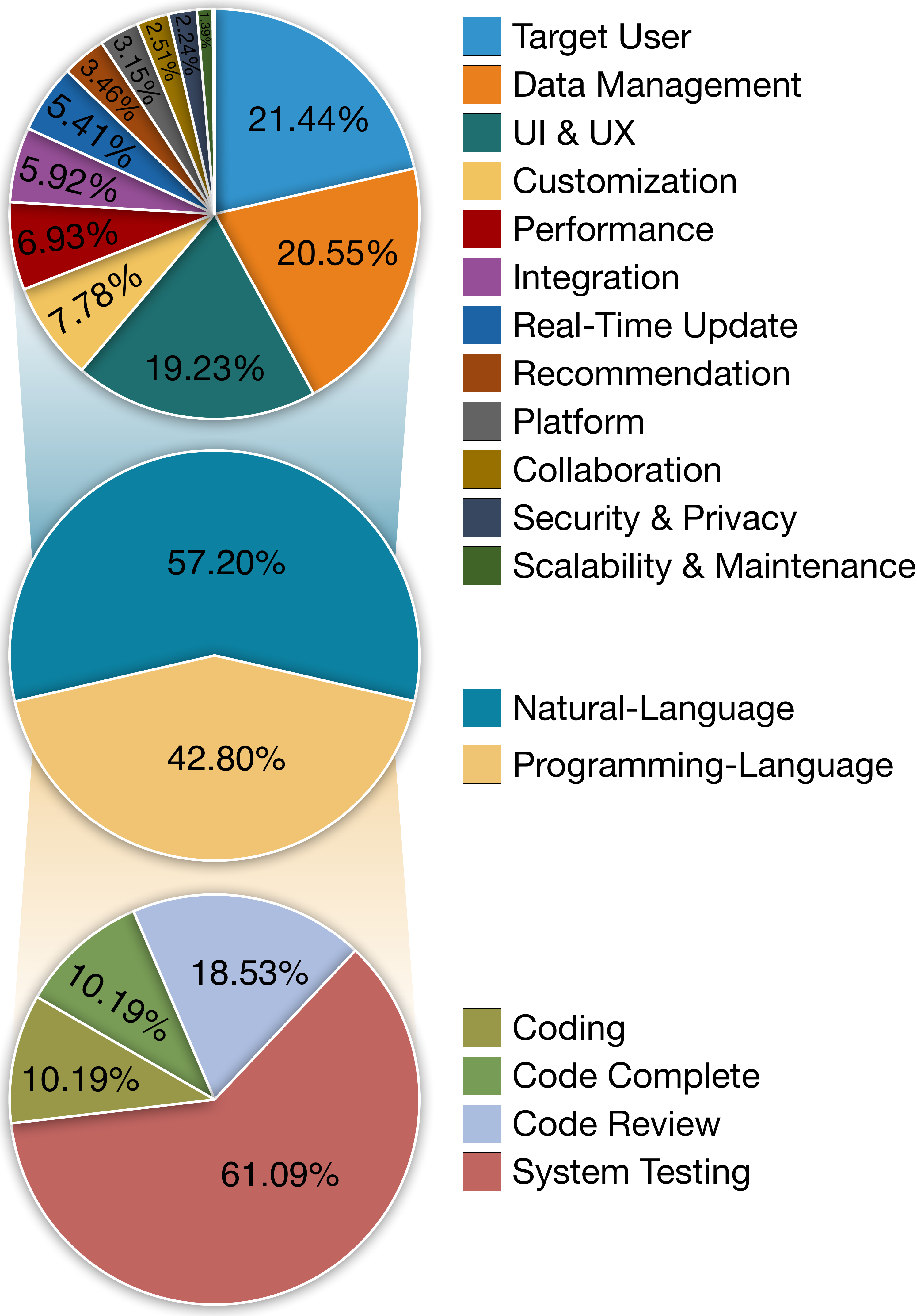}
    \caption{The utterance distribution of agent communications throughout the entire development process.}
    \label{fig:utterance_distribution}
\end{figure}

\begin{figure*}[t]
    \centering
    \includegraphics[width=0.99\textwidth]{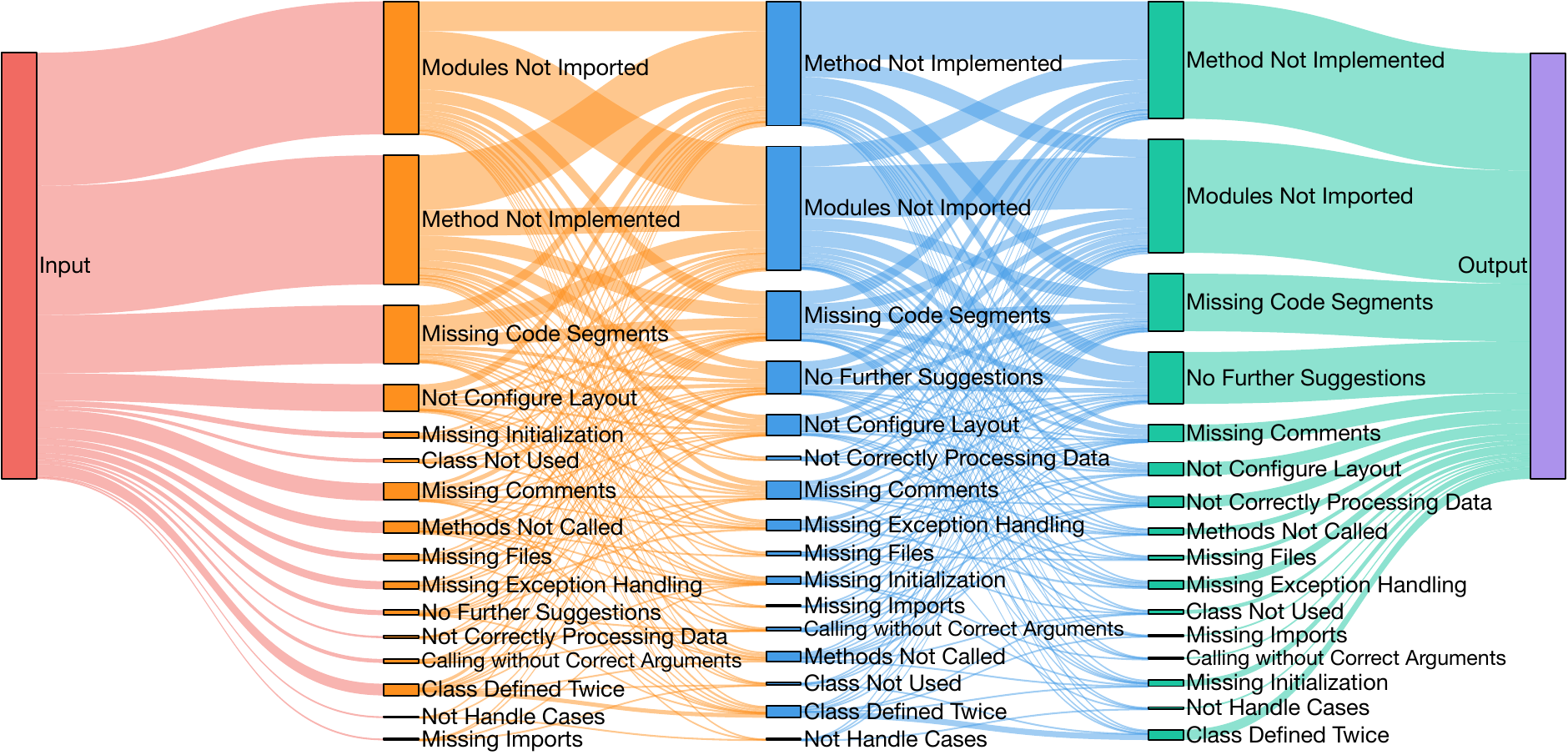}
    \caption{The chart demonstrates the distribution of suggestions made by a reviewer agent during a multi-round reviewing process, where each sector in the chart represents a different category of suggestion.}
    \label{fig:reviewer_sankey}
\end{figure*}

\begin{figure*}[t]
    \centering
    \includegraphics[width=0.99\textwidth]{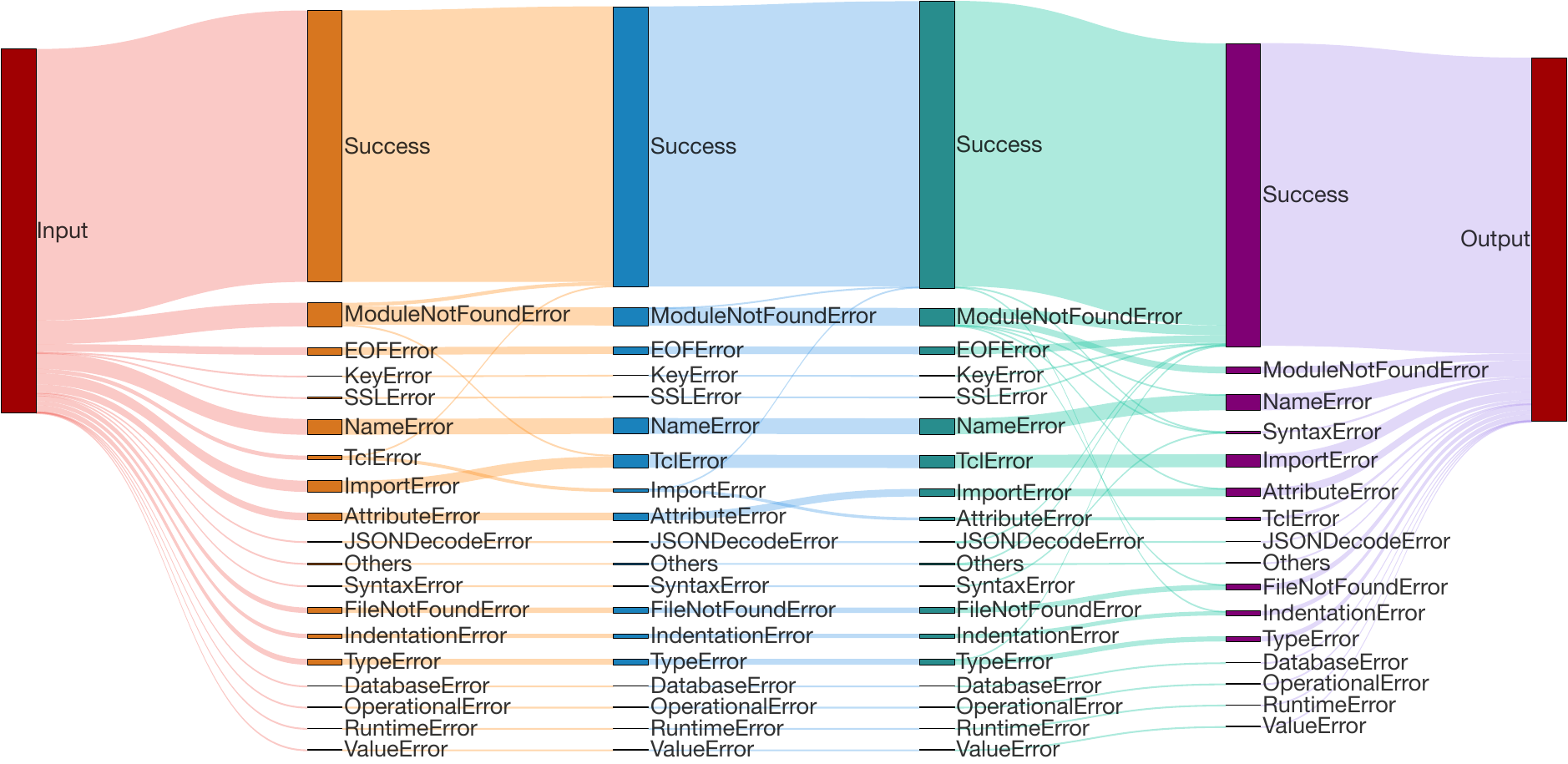}
    \caption{The diagram illustrates the progression of iterations in a multi-round testing process, where each colored column represents a dialogue round, showcasing the evolution of the solution through successive stages of testing.}
    \label{fig:test_sankey}
\end{figure*}

\subsection{Communication Analysis}

Our agent-driven software development paradigm promotes cooperative agents through effective communication for automated solution optimization. Phases in the chat chain have varying levels of engagement in natural and programming languages. We now analyze the content of their communications to understand linguistic effects.

Figure~\ref{fig:utterance_distribution} depicts a communication breakdown, with natural language at 57.20\%.
In the natural-language phase (\ie design), natural language communication plays a crucial role in the thorough design of the system, with agents autonomously discussing and designing aspects like target user, data management, and user interface.
Post-design phases show a balanced mix of coding, code completion, and testing activities, with most communication occurring during code reviews.
This trend is due to agents' self-reviews and code fixes consistently propelling software development; otherwise, progress halts when successive updates don't show significant changes, leading to a natural decrease in code review communications.

We explore the properties of static debugging dynamics in code reviews resulting from communication between reviewers and programmers, as depicted in Figure~\ref{fig:reviewer_sankey}.
The data uncovers that during the review phase, reviewers may spot different issues through language interactions. 
The programmer's intervention can transform certain issues into different ones or a state where no further suggestions are needed; the increasing proportion of the latter indicates successful software optimization.
Particularly, the "\texttt{Method Not Implemented}" issue is most common in communication between reviewers and programmers during code reviews, accounting for 34.85\% of discussions. This problem usually arises from unclear text requirements and the use of "placeholder" tags in Python code, necessitating additional manual adjustments.
Furthermore, the "\texttt{Module Not Imported}" issue often arises due to code generation omitting crucial details.
Apart from common problems, reviewers often focus on enhancing code robustness by identifying rare exceptions, unused classes, or potential infinite loops.

Likewise, we analyze the tester-programmer communication during the testing phase, illustrating the dynamic debugging dynamics in their multi-turn interactions with compiler feedback, as depicted in Figure~\ref{fig:test_sankey}.
The likelihood of successful compilation at each step is generally higher than encountering errors, with most errors persisting and a lower probability of transforming into different errors.
The most frequent error is "\texttt{ModuleNotFound}" (45.76\%), followed by "\texttt{NameError}" and "\texttt{ImportError}" (each at 15.25\%).
The observation highlights the model's tendency to overlook basic elements like an "import" statement, underscoring its difficulty in managing intricate details during code generation.
Besides, the tester also detects rarer errors like improperly initialized GUIs, incorrect method calls, missing file dependencies, and unused modules. 
The communicative dehallucination mechanism effectively resolves certain errors, frequently resulting in "compilation success" after code changes. There's a significantly low chance of returning to an error state from a successful compilation. Over time, the multi-turn communication process statistically shows a consistent decrease in errors, steadily moving towards successful software execution.

\section{Conclusion}
We have introduced ChatDev, an innovative multi-agent collaboration framework for software development that utilizes multiple LLM-powered agents to integrate fragmented phases of the waterfall model into a cohesive communication system. It features \textit{chat chain} organizing communication targets and \textit{dehallucination} for resolving coding hallucinations. 
The results demonstrate its superiority and highlight the benefits of multi-turn communications in software optimization.
We aim for the insights to advance LLM agents towards increased autonomy and illuminate the profound effects of "language" and its empowering role across an even broader spectrum of applications.

\section{Limitations}
Our study explores the potential of cooperative autonomous agents in software development, but certain limitations and risks must be considered by researchers and practitioners.
Firstly, the capabilities of autonomous agents in software production might be overestimated. While they enhance development quality, agents often implement simple logic, resulting in low information density. Without clear, detailed requirements, agents struggle to grasp task ideas. For instance, vague guidelines in developing a Snake game lead to basic representations; in information management systems, agents might retrieve static key-value placeholders instead of external databases. Therefore, it is crucial to clearly define detailed software requirements. Currently, these technologies are more suitable for prototype systems rather than complex real-world applications.
Secondly, unlike traditional function-level code generation, automating the evaluation of general-purpose software is highly complex. While some efforts have focused on \textit{Human Revision Cost}~\cite{hong2023metagpt}, manual verification for large datasets is impractical. Our paper emphasizes completeness, executability, consistency, and overall quality, but future research should consider additional factors such as functionalities, robustness, safety, and user-friendliness.
Thirdly, compared to single-agent approaches, multiple agents require more tokens and time, increasing computational demands and environmental impact. Future research should aim to enhance agent capabilities with fewer interactions.
Despite these limitations, we believe that engaging a broader, technically proficient audience can unlock additional potential directions in LLM-powered multi-agent collaboration.

\section*{Acknowledgments}
The work was supported by the National Key R\&D Program of China (No.2022ZD0116312), the Postdoctoral Fellowship Program of CPSF under Grant Number GZB20230348, and Tencent Rhino-Bird Focused Research Program.

\bibliography{references}

\end{document}